\documentclass[aps,twocolumn,preprintnumbers,superscriptaddress,prx]{revtex4-2}
\usepackage{graphicx}
\usepackage{amsmath}
\usepackage{amssymb}
\usepackage{color}
\usepackage{float}
    
\begin{document}

\title{Anomalous single-mode lasing induced by nonlinearity \\and the non-Hermitian skin effect}

\author{Bofeng Zhu}

\affiliation{Division of Physics and Applied Physics, School of Physical and Mathematical Sciences,\\
Nanyang Technological University, Singapore 637371, Singapore}

\affiliation{School of Electrical and Electronic Engineering, \\
Nanyang Technological University, Singapore 637371, Singapore}
  
\author{Qiang Wang}

\affiliation{Division of Physics and Applied Physics, School of Physical and Mathematical Sciences,\\
Nanyang Technological University, Singapore 637371, Singapore}

\author{Daniel Leykam}

\affiliation{Centre for Quantum Technologies, National University of Singapore, Singapore 117543, Singapore}

\author{Haoran Xue}

\affiliation{Division of Physics and Applied Physics, School of Physical and Mathematical Sciences,\\
Nanyang Technological University, Singapore 637371, Singapore}

\author{Qi Jie Wang}

\email{qjwang@ntu.edu.sg}

\affiliation{Division of Physics and Applied Physics, School of Physical and Mathematical Sciences,\\
Nanyang Technological University, Singapore 637371, Singapore}

\affiliation{School of Electrical and Electronic Engineering, \\
Nanyang Technological University, Singapore 637371, Singapore}
  
\affiliation{Centre for Disruptive Photonic Technologies, Nanyang Technological University, Singapore 637371, Singapore}

\author{Y.~D.~Chong}

\email{yidong@ntu.edu.sg}

\affiliation{Division of Physics and Applied Physics, School of Physical and Mathematical Sciences,\\
Nanyang Technological University, Singapore 637371, Singapore}

\affiliation{Centre for Disruptive Photonic Technologies, Nanyang Technological University, Singapore 637371, Singapore}

\begin{abstract}
  Single-mode operation is a desirable but elusive property for lasers operating at high pump powers.  Typically, single-mode lasing is attainable close to threshold, but increasing the pump power gives rise to multiple lasing peaks due to inter-modal gain competition.  We propose a laser with the opposite behavior: multi-mode lasing occurs at low output powers, but pumping beyond a certain value produces a single lasing mode, with all other candidate modes experiencing negative effective gain.  This phenomenon arises in a lattice of coupled optical resonators with non-fine-tuned asymmetric couplings, and is caused by an interaction between nonlinear gain saturation and the non-Hermitian skin effect.  The single-mode lasing is observed in both frequency domain and time domain simulations.  It is robust against on-site disorder, and scales up to large lattice sizes.  This finding might be useful for implementing high-power laser arrays.
\end{abstract}

\maketitle

Several unconventional laser systems have been invented, in the past few years, based on coupled-cavity configurations that achieve unusual effects not found in single-cavity designs.  These have drawn inspiration from a variety of other fields, such as non-Hermitian physics and condensed matter physics, and they include parity/time-reversal ($\mathcal{PT}$) symmetric lasers \cite{Feng2014, Hodaei2014, Qi2019}, lasers tuned by exceptional points \cite{Liertzer2012, Brandstetter2014, PartoReview2021}, topological lasers \cite{Zhao2018, Harari2018, Bandres2018, SmirnovaReview2020}, and, recently, lasers based on imaginary synthetic gauge fields \cite{Longhi2018, Longhi2018APL, Wong2021}.  A common objective of these efforts is to find lasers that sustain single-mode operation at high powers.  In conventional optical cavities, single-mode lasing occurs close to the lasing threshold, but increasing the pump causes more and more modes to acquire positive net gain.  The resulting multi-mode lasing is detrimental for applications requiring output beams that are spatially and temporally stable.  Strategies for maintaining single-mode operation have included built-in optical feedback \cite{Koyama1989}, free spectral range enlargement via sub-wavelength confinement \cite{Ma2011}, spatial structuring of the pump to match a specific mode \cite{Zhao2018, Liew2014}, and using topological edge modes to suppress mode localization \cite{Harari2018, Bandres2018}.

Some of the most interesting approaches to single-mode lasing have exploited the special features of non-Hermitian dynamics \cite{Qi2019, SmirnovaReview2020, Ozdemir2019, Kawabata2019, Ashida2020, PartoReview2021}.  Lasers are inherently non-Hermitian due to the presence of gain (pumping) and loss (outcoupling and dissipation), and some non-Hermitian systems can exhibit phenomena going far beyond simple mode amplification or damping.  For instance, $\mathcal{PT}$ symmetry breaking involves two modes of a non-Hermitian system coalescing and taking on identical frequencies and different gain/loss \cite{Ashida2020}; coupled-cavity lasers can use this to suppress half the modes that might lase \cite{Feng2014, Hodaei2014, Qi2019}, and for other forms of gain management \cite{Ozdemir2019}.  Other exotic aspects of non-Hermitian wave dynamics occur in periodic lattices
\cite{Rudner2009, Schomerus2013, Longhi2015, Longhi2015PRB, Malzard2015, Lee2016, Leykam2017, Gong2018, Brandenbourger2019}, like the ``non-Hermitian skin effect'' (NHSE), whereby the bulk modes of a non-Hermitian lattice collapse into boundary modes \cite{Hatano1996, Alvarez2018, Xiong2018, Yao2018, Zhang2020, Kawabata2020, Okuma2020, Ghatak2020, Helbig2020, Hofmann2020, Xiao2020, Claes2021, Zhang2021, Zhang2021HOSKIN, Weidemann2022}.  Longhi \cite{Longhi2018} has proposed a laser array based on a Su-Schrieffer-Heeger (SSH) lattice with asymmetric couplings that induce the NHSE.  By fine-tuning the couplings, the lattice can be made to host a single extended mode evolving from an SSH end mode.  This is made to lase via a $\mathcal{PT}$ symmetric pumping configuration, while all other modes are skin modes that receive little effective gain due to their spatial localization \cite{Longhi2018, Longhi2018APL, Wong2021}.

Most of these coupled-cavity laser designs have been based on linear lattice features applicable at or below the lasing threshold.  Above threshold, and especially at high powers, nonlinear effects become important but are usually detrimental to the intended functionality of a laser.  For instance, spatial hole burning, or the saturation of gain in regions where the field is most intense, tends to induce multi-mode lasing, since modes with profiles different from the lasing mode(s) acquire relatively higher gain \cite{Ge2010, Ge2014, Zhang2018}; it can also spoil useful symmetries such as $\mathcal{PT}$ symmetry.  One may ask whether nonlinearity could instead have a beneficial effect, say by interacting with the non-Hermitian properties to produce qualitatively new and useful behaviors in the above-threshold regime.

Here, we show that a coupled-cavity laser with non-fine-tuned asymmetric couplings can exhibit a behavior we call ``spontaneous single extended mode'' (SSEM) lasing, whereby increasing the pump above a certain level switches the laser from a multi-mode regime to a high power single-mode regime.  Under SSEM lasing, as the pump strength is raised to arbitrarily large values, the other modes experience decreasing effective gain, and recede further from threshold.  Although the mode suppression originates in NHSE-induced spatial localization, the SSEM in our model emerges spontaneously in the nonlinear, gain saturated regime.  The lattice sites are pumped uniformly, rather than spatially structuring the pump to select a specific mode \cite{Feng2014, Hodaei2014, Longhi2018}.  Moreover, the asymmetric couplings in the lattice need not be fine-tuned to produce an extended mode \cite{Longhi2018}.  The model can be realized in various ways, such as using a coupled resonator lattice with differential gain/loss on inter-site couplers \cite{Hafezi2011, Hafezi2013, Mittal2014, Harari2018, Longhi2015, Longhi2015PRB, Longhi2018, Zhu2020, Song2020}.  SSEM lasing could be used to implement laser arrays that lase in a single high-power mode, far above threshold.

\begin{figure}
  \centering
  \includegraphics[width=0.48\textwidth]{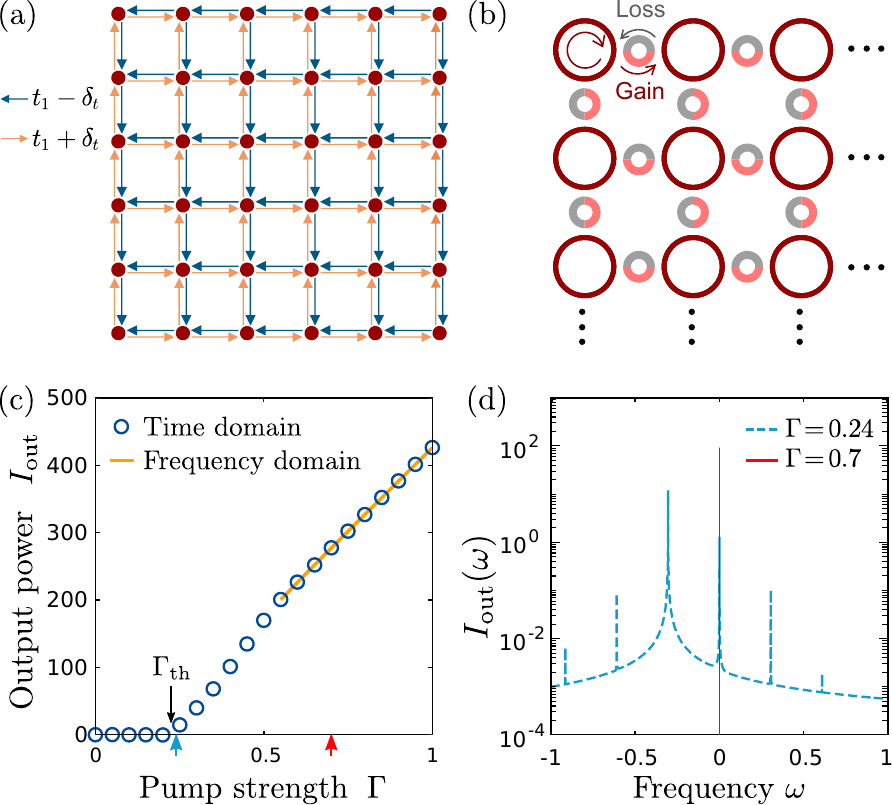}
  \caption{(a) Schematic of a 2D square lattice exhibiting spontaneous single extended mode (SSEM) lasing.  The sites (red circles) have uniformly-pumped saturable gain and linear loss, with asymmetric nearest neighbor couplings $t_1 \pm \delta_t$ (orange and blue arrows).  (b) Schematic of a possible realization based on a lattice of coupled ring resonators, with gain (pink) and loss (grey) on the coupling rings.  One circulation direction is assumed, indicated by the arrows.  (c) Output intensity $I_{\mathrm{out}}$ versus pump strength $\Gamma$ for a $6\times6$ lattice with $t_1=0.1$, $\delta_t=0.04$, and background loss $\gamma_0=0.2$.  Blue circles show time domain results, with initial conditions $\psi_{mn}(t=0) = (\alpha_{mn} + i \beta_{mn})f_0$, where $(m,n)$ is the site index, $\alpha_{mn}, \beta_{mn}$ are drawn independently from the standard normal distribution, and $f_0 = 0.01$ is a scale factor; $I_{\mathrm{out}}$ is obtained by averaging over an interval $t \in [20000,\,50000]$. The orange line shows the results of a self-consistent single-mode frequency domain calculation in the large-$\Gamma$ regime.  The threshold $\Gamma_{\mathrm{th}} = 0.2$ is indicated, while blue and red arrows mark the $\Gamma$ values used in the next subplot.  (d) Output spectrum for $\Gamma = 0.24$ (blue dashes) and $\Gamma = 0.7$ (red line), using the same lattice parameters as in (c).}
  \label{fig:schematic_latt}
\end{figure}

We consider the tight-binding model of Fig.~\ref{fig:schematic_latt}(a),
consisting of sites on a square lattice with one site per unit cell and asymmetric couplings $t_1 \pm \delta_t$ between nearest neighbors (orange and blue arrows).  This is a two-dimensional (2D) generalization of the Hatano-Nelson model \cite{Hatano1996}.  The Hamiltonian is $H = H_{0} + H_{1}$, where
\begin{multline} \label{finite_Ham}
  H_0 = \sum_{mn} \Big[ (t_1-\delta_t)(a_{mn}^{\dagger}a_{m+1,n} + a_{mn}^{\dagger}a_{m,n+1}) \\
  + (t_1+\delta_t)(a_{mn}^{\dagger}a_{m-1,n} + a_{mn}^{\dagger}a_{m,n-1}) \Big]
\end{multline}
is linear, non-Hermitian, and nonreciprocal; $a_{mn}$ denotes the annihilation operator on row $m$ and column $n$; and
\begin{equation}
  H_{1} = i \sum_{mn} \left[ \frac{\Gamma}{1+|\psi_{mn}|^2} - \gamma_0 \right]
  a_{mn}^\dagger a_{mn}
  \label{gainsat}
\end{equation}
describes on-site nonlinear gain and linear loss \cite{Ge2017, Harari2018}.  The parameter $\Gamma$ is the pump strength, $|\psi_{mn}|^2$ is the local intensity, and $\gamma_0$ is an on-site loss representing outcoupling and material absorption.  The pump and loss are spatially uniform, but the gain may be non-uniform due to $|\psi_{mn}|^2$ (i.e., gain saturation).  This tight-binding model can be realized with a lattice of coupled ring resonators \cite{Hafezi2011, Hafezi2013, Mittal2014, Harari2018}, as shown in Fig.~\ref{fig:schematic_latt}(b); each site is a ring resonator containing a laser medium, and the asymmetric couplings are implemented by placing unsaturated gain and loss on the arms of the coupling rings between neighboring sites \cite{Longhi2015, Longhi2015PRB, Longhi2018, Zhu2020, Song2020}.  Since such coupled-ring models are typically describable by tight-binding models \cite{Liang2013, Leykam2018}, we will focus on the latter.

We perform time domain simulations of the nonlinear Schr\"odinger equation $i\partial_t|\psi\rangle = H |\psi\rangle$, which has been widely applied in the investigation of laser dynamics \cite{Ge2017, Harari2018,Hassan2015,Ge2016,Teimourpour2016,Yang2020} (see Supplemental Materials \cite{SM}). The wavefunction $|\psi\rangle = [\psi_{11},\psi_{12}, \dots, \psi_{mn}, \dots]^T$ describes the amplitude and phase of the resonator mode \cite{Ge2016, Harari2018, Suh2004} at site $(m,n)$, where ${m,n} \in [1,2,...L]$ and $L$ is the number of sites on each side of square lattice. Fig.~\ref{fig:schematic_latt}(c) shows the resulting plot of output intensity $I_{\mathrm{out}}$ versus pump strength $\Gamma$, where $I_{\mathrm{out}} = \sum_{mn} |\psi_{mn}|^2$ (we assume equal outcoupling from each site, with normalized power units).  The model parameters are given in the figure caption; note that the dynamic range of $\Gamma$ is compatible with existing semiconductor lasers
\cite{Hodaei2014, Bandres2018, Zhao2018, Noda2014}. A lasing threshold occurs at $\Gamma_{\mathrm{th}} = 0.2$, above which the output power increases monotonically with $\Gamma$.  The output spectrum, plotted in Fig.~\ref{fig:schematic_latt}(d), contains multiple lasing peaks for weak pumping ($\Gamma = 0.24$), and a single peak under strong pumping ($\Gamma = 0.7$).  To verify the large-$\Gamma$ behavior, we perform a frequency domain calculation, using a nonlinear solver to find a self-consistent single-mode solution to the nonlinear time-independent Schr\"odinger equation (see Supplemental Materials \cite{SM}). This reproduces the $I_{\mathrm{out}}$ versus $\Gamma$ curve, as shown by the orange line in Fig.~\ref{fig:schematic_latt}(c).

\begin{figure}
  \centering
  \includegraphics[width=0.48\textwidth]{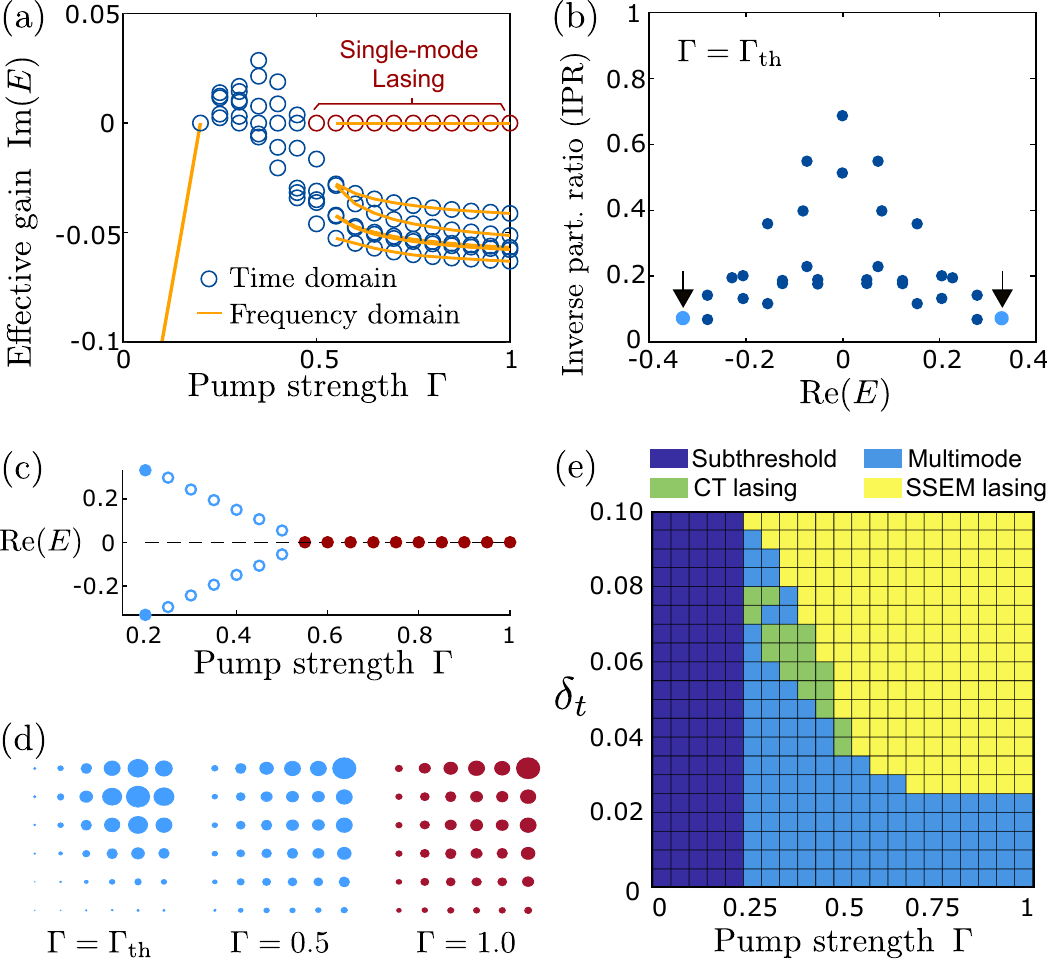}
  \caption{(a) Effective gain $\mathrm{Im}[E]$ versus pump strength $\Gamma$ for the various lattice eigenmodes.  Hollow circles are from the eigenenergies of the nonlinear Hamiltonians obtained via time domain simulations, using time-averaged intensities to calculate gain saturation.  Orange lines are frequency domain results obtained in the single-mode and below-threshold regimes.  (b) Inverse participation ratio (IPR), a measure of localization, for the eigenmodes at the lasing threshold $\Gamma_{\mathrm{th}}$.  The arrows indicate two modes at opposite band edges, which are related by $\mathcal{CT}$.  (c) $\mathrm{Re}(E)$ versus $\Gamma$ for the two modes indicated in (b).  With increasing $\Gamma$, they undergo a $\mathcal{CT}$ transition, and one of the resulting modes is the SSEM.  (d) Intensity distributions of the modes shown in (c) at different $\Gamma$.  The modes in each $\mathcal{CT}$ pair have the same intensities, so only one is plotted.  Like the original band edge modes, the SSEM is delocalized, even for large $\Gamma$.  (e) Phase diagram for the laser, indicating the below-threshold, multimode, and single-mode regimes, determined by time domain simulations for different values of the coupling asymmetry $\delta_t$ and pump strength $\Gamma$.  The single-mode regime is divided into SSEM lasing and $\mathcal{CT}$-broken lasing, as described in the text.  All other lattice parameters are the same as in Fig.~\ref{fig:schematic_latt}.}
  \label{fig:modal_gain}
\end{figure}

To help understand this phenomenon, we focus on two relevant features of the lattice.  First, in the linear regime and for $0 < \delta_t < t_1$, the 2D Hatano-Nelson model exhibits the NHSE \cite{Wong2021,Ezawa2022}, with skin modes localized to the upper-right corner of the lattice [Fig.~\ref{fig:schematic_latt}(a)].  Second, it can be shown that $\{H, \mathcal{CT}\} = 0$ for $\mathcal{C} = \sigma_{z} \otimes  \bold{I}_{M}$, where $\bold{I}_{M}$ is an identity matrix with rank $M=L^2/2$, $\sigma_z$ is the third Pauli matrix, and $\mathcal{T}$ is the time reversal (complex conjugation) operation \cite{Ge2017}.  The $\mathcal{CT}$ symmetry holds even in the nonlinear regime, provided the couplings are real and the diagonal terms of $H$---into which the saturable gain enters---are imaginary (see Supplemental Materials \cite{SM}).  Each eigenstate of $H$ is thus either $\mathcal{CT}$-symmetric with a purely imaginary eigenvalue, or $\mathcal{CT}$-broken with the partner eigenvalues related by $E_1=-E_2^{*}$ \cite{Ge2017}.

Figure~\ref{fig:modal_gain}(a) shows the evolution of the modal gain $\mathrm{Im}[E]$ with pump strength $\Gamma$ for the various lattice eigenmodes.  Below threshold, all modes have the same $\mathrm{Im}[E]$.  As $\Gamma$ increases, $\mathrm{Im}[E]$ increases linearly and reaches zero at the threshold $\Gamma_{\mathrm{th}}$.  The uniform pumping thus causes all modes to reach threshold simultaneously, unlike previous structured pumping schemes that selected, say, a $\mathcal{PT}$-broken mode \cite{Hodaei2014} or a topological mode \cite{Harari2018, Longhi2018}.  Above threshold, we calculate the modal gains [unfilled circles in Fig.~\ref{fig:modal_gain}(a)] from the effective \textit{nonlinear} Hamiltonian $H$ found by inserting the time-averaged intensities into Eq.~\eqref{gainsat}.  In the multimode regime, this replacement of the instantaneous intensity with a time average is an approximation, and results in the artifact that $\mathrm{Im}(E) > 0$ for some eigenmodes.  For larger $\Gamma$, however, the simulations settle into a time-independent intensity distribution, corresponding to a single lasing mode.  In this regime, one mode is found to have $\mathrm{Im}[E] = 0$, as expected, while all the others have $\mathrm{Im}[E] < 0$ and \textit{decreasing} with $\Gamma$.  In other words, the SSEM is an increasingly stable lasing mode as the pumping strength increases.  This behavior is corroborated by self-consistent frequency domain calculations, which yield modal gains exactly matching the time domain results in the SSEM lasing regime, as shown by the orange lines in Fig.~\ref{fig:modal_gain}(a).

By tracking the eigenmodes of the nonlinear Hamiltonians, we find that the SSEM originates from a pair of eigenmodes of the linear system, at opposite edges of the band.  These eigenmodes are $\mathcal{CT}$-broken partners and are poorly localized, as shown by the plot of the inverse participation ratio (IPR) in Fig.~\ref{fig:modal_gain}(b).  (The IPR, defined as $\sum_{mn} |\psi_{mn}|^4 / (\sum_{mn} |\psi_{mn}|^2)^2$, is a standard measure of localization  \cite{Thouless1974}).  As $\Gamma$ increases within the multi-mode regime, the two $\mathcal{CT}$-broken eigenmodes migrate across the band, eventually meeting at the center of the band and undergoing a $\mathcal{CT}$ transition, as shown in Fig.~\ref{fig:modal_gain}(c).  The evolution of the mode intensity distribution during this process is shown in Fig.~\ref{fig:modal_gain}(d).  After the transition, one of the $\mathcal{CT}$ symmetric modes becomes the SSEM, with eigenvalue pinned to $E = 0$, while its partner has $\mathrm{Im}(E) < 0$.  The SSEM has a lower IPR, or a larger mode area, than the remaining eigenmodes of the nonlinear Hamiltonian; in the Supplemental Materials, we show that its IPR scales inversely with the number of lattice sites, implying that it is a true extended mode \cite{SM}.  (Note that the presence of extended modes does not contradict the NHSE, which states that an extensive number of eigenmodes, but not necessarily all of them, become skin modes.)  All the other modes are more strongly localized, and with increasing $\Gamma$ have decreasing $\mathrm{Im}(E)$ as the SSEM saturates the gain throughout the lattice.

\begin{figure}
  \centering
  \includegraphics[width=0.48\textwidth]{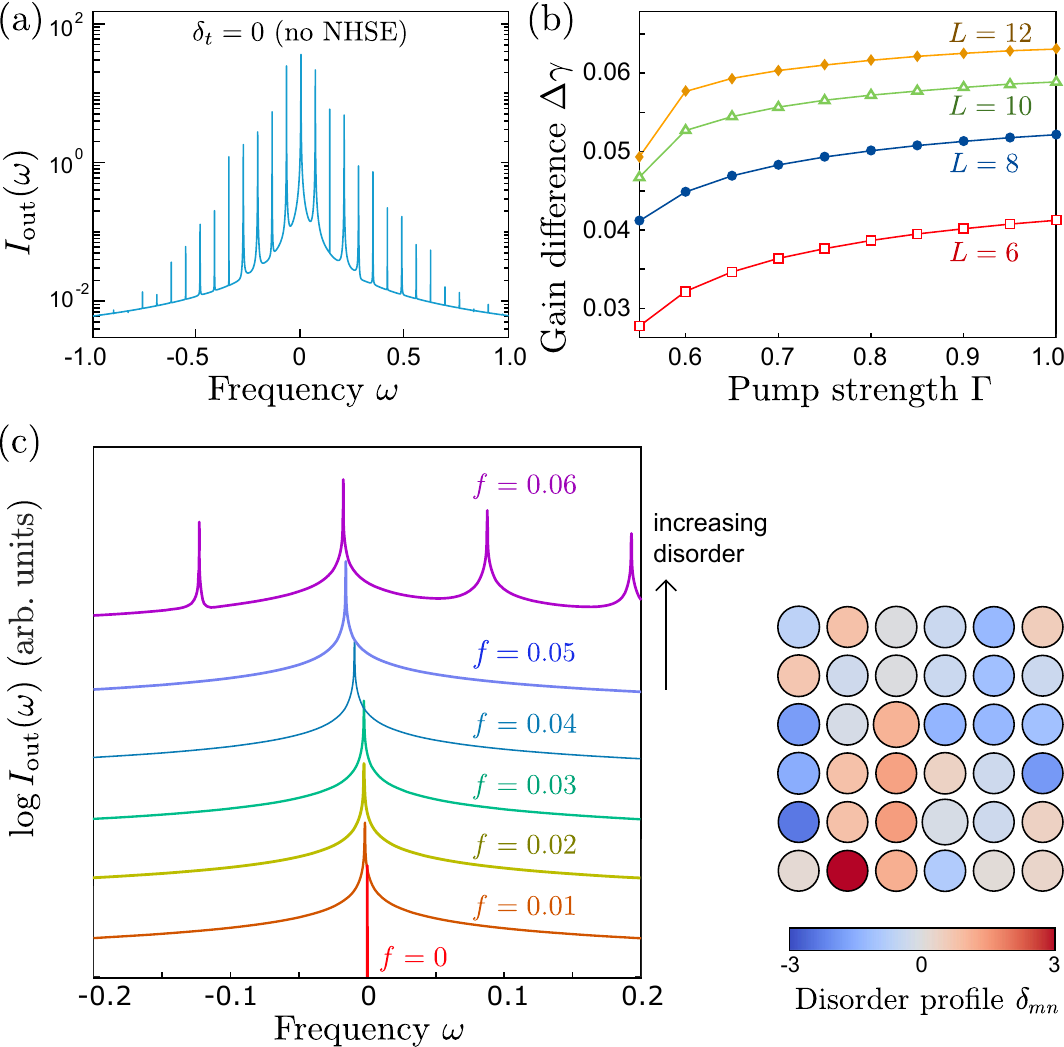}
  \caption{(a) Output spectrum for $\delta_t = 0$, obtained from a time domain simulation with lattice size $6\times6$, $\Gamma = 1$, and all other parameters the same as in Fig.~\ref{fig:schematic_latt}.  Since the couplings are symmetric, the NHSE is absent.  Multi-mode lasing is observed for all values of pump strength $\Gamma > \Gamma_{\mathrm{th}}$ we tested.  (b) Gain difference $\Delta\gamma$ between the single lasing mode and the next-highest-gain mode, plotted against pump strength $\Gamma$ in the SSEM lasing regime for lattices of varying size $L\times L$.  Larger $\Delta \gamma$ implies that the laser is further from the multi-mode regime.  Evidently, $\Delta \gamma$ increases with both $\Gamma$ and $L$, and its increase with $\Gamma$, for fixed $L$, is also consistent with Fig.~\ref{fig:modal_gain}(a).  (c) Effects of on-site disorder.  Left panel: output spectra, plotted with different vertical offsets, in lattices with real on-site detunings $f \cdot \delta_{mn}$, where $f$ is a scale factor and $\delta_{mn}$ is drawn independently from the standard normal distribution for each site.  The lattice is pumped at $\Gamma = 1$, with all other parameters the same as in Fig.~\ref{fig:schematic_latt}.  Results are shown for several values of $f$, using the same $\delta_{mn}$ distribution for each case.  For sufficiently weak disorder, single-mode lasing is observed, though the lasing mode is not pinned to zero.  For strong disorder, multi-mode lasing occurs.  Right panel: the random disorder profile $\delta_{mn}$ used to generate these results. }
  \label{fig:size}
\end{figure}

Figure~\ref{fig:modal_gain}(e) shows the phase diagram of the laser, derived from time domain simulations and plotted against the coupling asymmetry $\delta_t$ and the pump strength $\Gamma$.  For small $\delta_t$, we only observe multi-mode lasing since the NHSE is weak.  This is exemplified by the strongly multi-modal spectrum plotted in Fig.~\ref{fig:size}(a), obtained with $\delta_t = 0$ (i.e., symmetric couplings) for which the NHSE is absent.  For larger $\delta_t$, single-mode lasing always sets in at high pump strengths.  The threshold for the SSEM regime decreases with $\delta_t$, and approaches the lasing threshold as $\delta_t \rightarrow t_1$.  There is also an intermediate regime, indicated in green in the phase diagram, which we call ``$\mathcal{CT}$-broken lasing''.  This occurs when the two aforementioned $\mathcal{CT}$-broken modes have real frequencies $\pm E_0$ but have not yet reached their $\mathcal{CT}$ transition, while all other modes have $\mathrm{Im}(E) < 0$.  Since the two modes have identical spatial intensity profiles, they compete strongly with each other, resulting in lasing at either $+E_0$ or $-E_0$ (chosen spontaneously based on the initial conditions).  This persists over a relatively small range of $\Gamma$; with stronger pumping, the $\mathcal{CT}$ transition occurs and the SSEM emerges as the sole lasing mode at $E = 0$.

The results presented above were obtained with $6\times 6$ lattices, but SSEM lasing is also found to occur in larger lattices, as well as in one-dimensional (1D) lattices (see Supplemental Materials \cite{SM}).  Figure~\ref{fig:size}(b) plots the gain difference $\Delta \gamma = \mathrm{Im}(E_0) - \mathrm{Im}(E_1)$ versus $\Gamma$ in the SSEM lasing regime, where $E_0 = 0$ is the energy of the lasing mode and $E_1$ is the energy of the mode with the next-highest gain, for 2D lattices of size $L\times L$.  The increase of $\Delta \gamma$ with $\Gamma$ (i.e., the other modes receding further from threshold) is consistent with the $6\times6$ results shown in Fig.~\ref{fig:modal_gain}(a).  Moreover, for each $\Gamma$ we find that $\Delta \gamma$ increases with $L$.  This indicates that the SSEM lasing phenomenon can scale to large lattices to achieve high-power lasing.  In the Supplemental Materials, we show that SSEM lasing can also be observed in a 1D lattice (the Hatano-Nelson model \cite{Hatano1996} with additional on-site nonlinear gain and linear loss).  The main difference is that 1D lattices have larger mode spacings, so the occurrence of single-mode lasing may not be as striking as in 2D \cite{SM}.

The gain difference between lasing and non-lasing modes, which increases with pump strength and system size, provides a measure of robustness against disorder.  In Fig.~\ref{fig:size}(c), we plot the output spectra for $6\times6$ lattices with additional real on-site mass terms $f \cdot \delta_{mn}$, where $f$ is a scale factor and $\delta_{mn}$ is drawn independently from the standard normal distribution for each site.  This mimics random frequency detunings induced by fabrication defects or other sources of disorder.  At fixed pump strength $\Gamma$, the system exhibits single-mode lasing over a range of $f > 0$, though the lasing mode is no longer pinned to zero since the $\mathcal{CT}$ symmetry is spoiled by the real on-site terms \cite{SM}.  Multimode lasing sets in if the disorder is sufficiently strong (in this case, $f \gtrsim 0.06$; this should scale with $\Delta \gamma$).  Another reason for the robustness may be that disorder-induced mode localization is counteracted by the asymmetric couplings, as suggested by previous studies \cite{Hatano1996, Longhi2015, Gong2018}.  Although the present behavior may be reminiscent of laser mode stabilization induced by linear $\mathcal{PT}$ symmetry \cite{Feng2014, Hodaei2014, Qi2019} or $\mathcal{CT}$ symmetry \cite{Ge2017, Hentinger2022}, the underlying mechanism is different, since SSEM lasing arises in the nonlinear gain saturated regime and its stability improves with system size. 

In summary, we predict that an unusual form of single-mode lasing arises in lattices with asymmetric couplings: the laser can be multi-mode for weak pumping, but single-mode at high output powers.  Strikingly, the stability of the single lasing mode increases with both pump strength and lattice size.  The possibility of exotic phenomena emerging from the interplay of non-Hermiticity and nonlinearity has been explored in previous works; for instance, coupled cavities operating near an exceptional point have been shown to exhibit an inverted power curve, with stronger pumping lowering the output power \cite{Liertzer2012, Brandstetter2014}.  There have been proposals to use the non-Hermitian skin effect (NHSE) for single-mode lasing \cite{Longhi2018, Longhi2018APL, Wong2021}, but our system is different as the special lasing mode emerges in the gain saturated regime and is not spatially selected by the pump (our lattice is pumped uniformly).  Other intriguing consequences of combining the NHSE with Kerr nonlinearity, rather than gain saturation, have also recently been studied \cite{Yuce2021, Ezawa2022, Lang2021}.  To realize our model, a promising approach may be to use an array of coupled ring resonators with differential gain and/or loss on the coupling rings to induce asymmetric couplings between site rings \cite{Leykam2018, SM}, fabricated on a gain medium with a large dynamic range \cite{Faist2013, Noda2014}.

This work was supported by the Singapore MOE Academic Research Fund Tier 3 Grant MOE2016-T3-1-006, Tier 2 Grant MOE2019-T2-2-085, Tier 1 Grants RG148/20, and Singapore National Research Foundation (NRF) Competitive Research Program (CRP) NRF-CRP18-2017-02 and NRF-CRP23-2019-0007.


\begin{thebibliography}{99}


\bibitem{Feng2014}     
  L.~Feng, Z.~J.~Wong, R.~M.~Ma, Y.~Wang, and X.~Zhang,
  Single-mode laser by parity-time symmetry breaking,
  Science~\textbf{346}, 972 (2014).  

\bibitem{Hodaei2014}     
  H.~Hodaei, M.~A.~Miri, M.~Heinrich, D.~N.~Christodoulides, and M.~Khajavikhan,
  Parity-time–symmetric microring lasers,
  Science~\textbf{346}, 975 (2014).

\bibitem{Qi2019}
  B.~Qi, H.~Chen, L.~Ge, P.~Berini, and R.~Ma,
  Parity-Time Symmetry Synthetic Lasers: Physics and Devices,  
  Adv.~Opt.~Mater.~\textbf{7}, 1900694 (2019). 

  
\bibitem{Liertzer2012}
  M.~Liertzer, L.~Ge, A.~Cerjan, A.~D.~Stone, H.~E.~T{\"u}reci, and S.~Rotter,
  Pump-Induced Exceptional Points in Lasers,
  Phys.~Rev.~Lett.~\textbf{108}, 173901 (2012). 

\bibitem{Brandstetter2014}
  M.~Brandstetter, M.~Liertzer, C.~Deutsch, P.~Klang, J.~Sch{\"o}berl, H.~E.~T{\"u}reci, G.~ Strasser, K.~Unterrainer, and S.~Rotter,
  Reversing the pump dependence of a laser at an exceptional point,
  Nat.~Commun.~\textbf{5}, 4034 (2014).

\bibitem{PartoReview2021}
  M.~Parto, Y.~Liu, B.~Bahari, M.~Khajavikhan, and D.~Christodoulides,
  Non-Hermitian and topological photonics: Optics at an exceptional point,
  Nanophotonics~\textbf{10}, 403 (2021).

  
\bibitem{Zhao2018}
  H.~Zhao, P.~Miao, M.~H.~Teimourpour, S.~Malzard, R.~E.~Ganainy, H.~Schomerus, and L.~Feng,
  Topological hybrid silicon microlasers,
  Nat.~Commun.~\textbf{9}, 981 (2018).  

\bibitem{Harari2018}
  G.~Harari, Miguel.~A.~Bandres, Y.~Lumer, M.~C.~Rechtsman, Y.~D.~Chong, M.~Khajavikhan,
  D.~N.~Christodoulides, and M.~Segev,
  Topological insulator laser: Theory,
  Science~\textbf{359}, eaar4003 (2018).

\bibitem{Bandres2018}
  M.~A.~Bandres, S.~Wittek, G.~Harari, M.~Parto, J.~Ren, M.~Segev, D.~N.~Christodoulides, and M.~Khajavikhan,
  Topological insulator laser: Experiments,
  Science~\textbf{359}, eaar4005 (2018).

\bibitem{SmirnovaReview2020}
  D.~Smirnova, D.~Leykam, Y.~D.~Chong and Y.~Kivshar,
  Nonlinear topological photonics,
  Appl.~Phys.~Rev.~\textbf{7}, 021306 (2020).  


\bibitem{Longhi2018}
  S.~Longhi,
  Non-Hermitian gauged topological laser arrays,
  Ann.~Phys.~\textbf{530}, 1800023 (2018). 

\bibitem{Longhi2018APL}
  S.~Longhi and L.~Feng,
  Mitigation of dynamical instabilities in laser arrays via non-Hermitian coupling,
  APL.~Photonics~\textbf{3}, 060802 (2018). 

\bibitem{Wong2021} 
  S.~Wong and S.~S.~Oh,
  Topological bulk lasing modes using an imaginary gauge field,
  Phys.~Rev.~Research~\textbf{3}, 033042 (2021).


\bibitem{Koyama1989}
  F.~Koyama, S.~Kinoshita, and K.~Iga,
  Room-temperature continuous wave lasing characteristics of a GaAs vertical cavity surface-emitting laser,
  Appl.~Phys.~Lett.~\textbf{55}, 221 (1989).  

\bibitem{Ma2011}
  M.~R.~Ma, R.~F.~Oulton, V.~J.~Sorger, G.~Bartal, and X.~Zhang,
  Room-temperature sub-diffraction-limited plasmon laser by total internal reflection,
  Nat.~Mater.~\textbf{10}, 110 (2011).

\bibitem{Liew2014}
  S.~F.~Liew, B.~Redding, L.~Ge, G.~S.~Solomon, and H.~Cao,
  Active control of emission directionality of semiconductor microdisk lasers,
  Appl.~Phys.~Lett.~\textbf{104}, 231108 (2014).  


\bibitem{Ozdemir2019}
  {\c{S}}.~K.~{\"O}zdemir, S.~Rotter, F.~Nori, and L.~Yang,
  Parity-time symmetry and exceptional points in photonics,
  Nat.~Mater.~\textbf{18}, 783 (2019).  

\bibitem{Kawabata2019}
  K.~Kawabata, K.~Shiozaki, M.~Ueda, and M.~Sato, 
  Symmetry and Topology in Non-Hermitian Physics,
  Phys.~Rev.~X~\textbf{9}, 041015 (2019).  

\bibitem{Ashida2020}
  Y.~Ashida, Z.~Gong, and M.~Ueda,
  Non-Hermitian physics,
  Adv.~Phys.~\textbf{69}, 249 (2020).


 
\bibitem{Rudner2009}
 M.~S.~Rudner and L.~S.~Levitov,
 Topological Transition in a Non-Hermitian Quantum Walk,
 Phys.~Rev.~Lett.~\textbf{102}, 065703 (2009).

\bibitem{Schomerus2013}
  H.~Schomerus,
  Topologically protected midgap states in complex photonic lattices,
  Opt.~Lett.~\textbf{38}, 1912 (2013).

\bibitem{Longhi2015} 
  S.~Longhi, D.~Gatti, and G.~D.~Valle,
  Robust light transport in non-Hermitian photonic lattices,
  Sci.~Rep.~\textbf{5}, 13376 (2015).
  
\bibitem{Longhi2015PRB}
  S.~Longhi, D.~Gatti, and G.~D.~Valle,
  Non-Hermitian transparency and one-way transport in low-dimensional lattices by an imaginary gauge field,
  Phys.~Rev.~B~\textbf{92}, 094204 (2015).  
  
\bibitem{Malzard2015}
 S.~Malzard, C.~Poli, and H.~Schomerus,
 Topologically Protected Defect States in Open Photonic Systems with Non-Hermitian Charge-Conjugation and Parity-Time Symmetry,
 Phys.~Rev.~Lett.~\textbf{115}, 200402 (2015).

\bibitem{Lee2016}
 T.~E.~Lee,
 Anomalous Edge State in a Non-Hermitian Lattice,
 Phys.~Rev.~Lett.~\textbf{116}, 133903 (2016).  
  
\bibitem{Leykam2017}
 D.~Leykam, K.~Y.~Bliokh, C.~Huang, Y.~D.~Chong, and F.~Nori,
 Edge Modes, Degeneracies, and Topological Numbers in Non-Hermitian Systems,
 Phys.~Rev.~Lett.~\textbf{118}, 040401 (2017).
  
\bibitem{Gong2018}
  Z.~Gong, Y.~Ashida, K.~Kawabata, K.~Takasan, S.~Higashikawa, and M.~Ueda,
  Topological Phases of Non-Hermitian Systems,
  Phys.~Rev.~X~\textbf{8}, 031079 (2018).   

\bibitem{Brandenbourger2019} 
  M.~Brandenbourger, X.~Locsin, E.~Lerner, and C.~Coulais,
  Non-reciprocal robotic metamaterials,
  Nat.~Commun.~\textbf{10}, 4608 (2019).  


\bibitem{Hatano1996}
  N.~Hatano and D.~R.~Nelson,
  Localization Transitions in Non-Hermitian Quantum Mechanics,
  Phys.~Rev.~Lett.~\textbf{77}, 570 (1996).

\bibitem{Alvarez2018}
  V.~M.~Martinez Alvarez, J.~E.~Barrios Vargas, and L.~E.~F.~Foa Torres,
  Non-Hermitian robust edge states in one dimension: Anomalous localization and eigenspace condensation at exceptional points,
  Phys.~Rev.~B~\textbf{97}, 121401(R) (2018).

\bibitem{Xiong2018}
  Y.~Xiong,
  Why does bulk boundary correspondence fail in some non-Hermitian topological models,
  J.~Phys.~Commun.~\textbf{2}, 035043 (2018).

\bibitem{Yao2018}
  S.~Yao and Z.~Wang,
  Edge States and Topological Invariants of Non-Hermitian Systems,
  Phys.~Rev.~Lett.~\textbf{121}, 086803 (2018).

\bibitem{Zhang2020}
  K.~Zhang, Z.~Yang, and C.~Fang, 
  Correspondence between Winding Numbers and Skin Modes in Non-Hermitian Systems,
  Phys.~Rev.~Lett.~\textbf{125}, 126402 (2020). 

\bibitem{Kawabata2020}
  K.~Kawabata, M.~Sato, and K.~Shiozaki,
  Higher-order non-Hermitian skin effect,
  Phys.~Rev.~B~\textbf{102}, 205118 (2020).  

\bibitem{Okuma2020}
  N.~Okuma, K.~Kawabata, K.~Shiozaki, and M.~Sato, 
  Topological Origin of Non-Hermitian Skin Effects,
  Phys.~Rev.~Lett.~\textbf{124}, 086801 (2020).

\bibitem{Ghatak2020} 
  A.~Ghatak, M.~Brandenbourger, J.~van~Wezel, and C.~Coulais,
  Observation of non-Hermitian topology and its bulk-edge correspondence in an active mechanical metamaterial,
  Proc.~Natl.~Acad.~Sci.~U.S.A.~\textbf{117}, 29561 (2020).  

\bibitem{Helbig2020}
  T.~Helbig, T.~Hofmann, S.~Imhof, M.~Abdelghany, T.~Kiessling, L.~W.~Molenkamp, C.~H.~Lee, A.~Szameit, M.~Greiter, and R.~Thomale,
  Generalized bulk-boundary correspondence in non-Hermitian topolectrical circuits,
  Nat.~Phys.~\textbf{16}, 747 (2020).

\bibitem{Hofmann2020} 
  T.~Hofmann, T.~Helbig, F.~Schindler, N.~Salgo, M.~Brzezi{\'n}ska, M.~Greiter, T.~Kiessling, D.~Wolf, A.~Vollhardt, A.~Kaba{\v{s}}i, C.~H.~Lee, A.~Bilu{\v{s}}i´c, R.~Thomale, and T.~Neupert,
  Reciprocal skin effect and its realization in a topolectrical circuit,
  Phys.~Rev.~Research~\textbf{2}, 023265 (2020).

\bibitem{Xiao2020}
  L.~Xiao, T.~Deng, K.~Wang, G.~Zhu, Z.~Wang, W.~Yi, and P.~Xue,
  Non-Hermitian bulk-boundary correspondence in quantum dynamics,
  Nat.~Phys.~\textbf{16}, 761 (2020).

\bibitem{Claes2021}
  J.~Claes and T.~L.~Hughes,
  Skin effect and winding number in disordered non-Hermitian systems,
  Phys.~Rev.~B~\textbf{103}, L140201 (2021).  

\bibitem{Zhang2021}
  K.~Zhang, Z.~Yang, and C.~Fang,
  Universal non-Hermitian skin effect in two and higher dimensions,
  Nat.~Commun.~\textbf{13}, 2496 (2022). 

\bibitem{Zhang2021HOSKIN} 
  X.~Zhang, Y.~Tian, J.~Jiang, M.~Lu, and Y.~Chen,
  Observation of higher-order non-Hermitian skin effect,
  Nat.~Commun.~\textbf{12}, 5377 (2021).  
  
\bibitem{Weidemann2022}
 S.~Weidemann, M.~Kremer, S.~Longhi, and A.~Szameit,
 Topological triple phase transition in non-Hermitian Floquet quasicrystals,
 Nature~\textbf{601}, 354 (2022).
 

\bibitem{Ge2010}
  L.~Ge, Y.~D.~Chong, and A.~D.~Stone,
  Steady-state \textit{ab initio} laser theory: Generalizations and analytic results,
  Phys. Rev. A 82, 063824 (2010).  

\bibitem{Ge2014}
  L.~Ge, O.~Malik, and H.~E.~T\"ureci,
  Enhancement of laser power-efficiency by control of spatial hole burning interactions,
  Nat.~Photonics~\textbf{8}, 871 (2014). 

\bibitem{Zhang2018}
  Z.~Zhang, P.~Miao, J.~Sun, S.~Longhi, N.~M.~Litchinitser, and L.~Feng,
  Elimination of Spatial Hole Burning in Microlasers for Stability and Efficiency Enhancement,
  ACS~Photonics~\textbf{5}, 3016 (2018).  
 

\bibitem{Hafezi2011}
  M.~Hafezi, E.~A.~Demler, M.~D.~Lukin, and J.~M.~Taylor,
  Robust optical delay lines with topological protection,
  Nat.~Phys.~\textbf{7}, 907 (2011).
  
\bibitem{Hafezi2013}
  M.~Hafezi, S.~Mittal, J.~Fan, A.~Migdall, and J.~M.~Taylor,
  Imaging topological edge states in silicon photonics,
  Nat.~Photonics~\textbf{7}, 1001 (2013).

\bibitem{Mittal2014}
  S.~Mittal, J.~Fan, S.~Faez, A.~Migdall, J.~M.~Taylor, and M.~Hafezi,
  Topologically Robust Transport of Photons in a Synthetic Gauge Field,
  Phys.~Rev.~Lett.~\textbf{113}, 087403 (2014). 


\bibitem{Zhu2020}
  X.~Zhu, H.~Wang, S.~K.~Gupta, H.~Zhang, B.~Xie, M.~Lu, and Y.~Chen,
  Photonic non-Hermitian skin effect and non-Bloch bulk-boundary correspondence,
  Phys.~Rev.~Research~\textbf{2}, 013280 (2020).  
  
\bibitem{Song2020}
  Y.~Song, W.~Liu, L.~Zheng, Y.~Zhang, B.~Wang , and P.~Lu,
  Two-Dimensional Non-Hermitian Skin Effect in a Synthetic Photonic Lattice,
  Phys.~Rev.~Applied~\textbf{14}, 064076 (2020).  


\bibitem{Ge2017}  
  L.~Ge,
  Symmetry-protected zero-mode laser with a tunable spatial profile,
  Phys.~Rev.~A~\textbf{95}, 023812 (2017).


\bibitem{Liang2013}
  G.~Q.~Liang and Y.~D.~Chong,
  Optical Resonator Analog of a Two-Dimensional Topological Insulator,
  Phys.~Rev.~Lett.~\textbf{110}, 203904 (2013). 

\bibitem{Leykam2018}
  D.~Leykam, S.~Mittal, M.~Hafezi, and Y.~D.~Chong,
  Reconfigurable Topological Phases in Next-Nearest-Neighbor Coupled Resonator Lattices,
  Phys.~Rev.~Lett.~\textbf{121}, 023901 (2018). 


\bibitem{SM}
  \textcolor{blue}{See online Supplemental Materials [...] for the lattice symmetries analyses; details of coupled-ring lattice model, time-domain and frequency-domain analyses; lasing in various 1D and 2D lattices, which includes Ref. \cite{Zeng2020}}
  

\bibitem{Hassan2015}
  A.~U.~Hassan, H.~Hodaei, M.~A.~Miri, M.~Khajavikhan, and D.~N.~Christodoulides,
  Nonlinear reversal of the $\mathcal{PT}$-symmetric phase transition in a system of coupled semiconductor microring resonators,
  Phys.~Rev.~A \textbf{92}, 063807 (2015).
  
\bibitem{Ge2016}
  L.~Ge and R.~El-Ganainy,
  Nonlinear modal interactions in parity-time ($\mathcal{PT}$) symmetric lasers,
  Sci.~Rep.~\textbf{6}, 24889 (2016).
  
\bibitem{Teimourpour2016}
  M.~H.~Teimourpour, L.~Ge, D.~N. Christodoulides, and R.~El-Ganainy,
  Non-Hermitian engineering of single mode two dimensional laser arrays,
  Sci.~Rep.~\textbf{6}, 33253 (2016).

\bibitem{Yang2020}   
  Z.~Yang, E.~Lustig, G.~Harari, Y.~Plotnik, Y.~Lumer, M.~A.~Bandres and M.~Segev,
  Mode-Locked Topological Insulator Laser Utilizing Synthetic Dimensions,
  Phys.~Rev.~X~\textbf{10}, 011059 (2020)
  

\bibitem{Suh2004}
  W.~Suh, Z,~Wang and S.~Fan, Temporal coupled-mode theory and the presence of non-orthogonal modes in lossless multimode cavities, IEEE J.~Quantum~Electron.~\textbf{40}, 1511 (2004).


\bibitem{Noda2014}
  K.~Hirose, Y.~Liang, Y.~Kurosaka, A.~Watanabe, T.~Sugiyama, and S.~Noda, Watt-class high-power, high-beam-quality photonic-crystal lasers, Nat.~Photonics~\textbf{8}, 406 (2014).


\bibitem{Ezawa2022}      
  M.~Ezawa,
  Dynamical nonlinear higher-order non-Hermitian skin effects and topological trap-skin phase,
  Phys.~Rev.~B~\textbf{105}, 125421 (2022).  


\bibitem{Thouless1974}
  D.~J.~Thouless, Electrons in disordered systems and the theory of localization, Phys.~Rep.~13, 93 (1974).

\bibitem{Hentinger2022}
  F.~Hentinger, M.~Hedir, B.~Garbin, M.~Marconi, L.~Ge, F.~Raineri, J.~A.~Levenson, and A.~M.~Yacomotti,
  Direct observation of zero modes in a non-Hermitian optical nanocavity array,
  Photonics~Res.~\textbf{10}, 574 (2022).  
  

\bibitem{Yuce2021}  
  C.~Yuce,
  Nonlinear non-Hermitian skin effect,
  Phys.~Lett.~A~\textbf{408}, 127484 (2021).
  
\bibitem{Lang2021}
  L.-J.~Lang, S.-L.~Zhu, and Y.~D.~Chong,
  Non-Hermitian topological end breathers,
  Phys.~Rev.~B \textbf{104}, L020303 (2021).



\bibitem{Faist2013}
  J.~Faist,
  Quantum Cascade Lasers,~1st ed. (Oxford University Press, Oxford, 2013),~Chap.~7.    

  

 
\bibitem{Zeng2020}
  Y.~Zeng, U.~Chattopadhyay, B.~Zhu, B.~Qiang, J.~ Li, Y.~Jin, L.~Li, A.~G.~Davies, E.~H.~Linfield, B.~Zhang, Y.~D.~Chong and Q.~J.~Wang,
  Electrically pumped topological laser with valley edge modes,
  Nature~\textbf{578}, 246 (2020).  
     
\end{thebibliography}
\end{document}